\documentclass[aps,prl,reprint,amsmath,amssymb]{revtex4-1}

\makeatletter
\let\frontmatter@footnote@produce\frontmatter@footnote@produce@endnote
\makeatother

\usepackage{graphicx}
\usepackage{physics}
\usepackage{bm}  
\usepackage{amsmath} 
\usepackage{braket} 
\usepackage[colorlinks,allcolors=blue,urlcolor=blue]{hyperref}

\begin{document}

\title{Elliott-Yafet Spin-Phonon Relaxation Times from First Principles}

\author{Jinsoo Park}
\author{Jin-Jian Zhou}%
\author{Marco Bernardi}
\email{bmarco@caltech.edu}
\affiliation{Department of Applied Physics and Materials Science, California Institute of Technology, Pasadena, California 91125}
\date{\today}
%
\begin{abstract}
We present a first-principles approach for computing the phonon-limited $T_1$ spin relaxation time due to the Elliot-Yafet mechanism. 
Our scheme combines fully-relativistic spin-flip electron-phonon interactions with an approach to compute the effective spin of band electrons in materials with inversion symmetry. 
We apply our method to silicon and diamond, for which we compute the temperature dependence of the spin relaxation times and analyze the contributions to spin relaxation from different phonons and valley processes. 
The computed spin relaxation times in silicon are in excellent agreement with experiment in the 50$-$300~K temperature range. In diamond, we predict intrinsic spin relaxation times of 540~$\mu$s at 77~K and 2.3~$\mu$s at 300~K. 
Our work enables precise predictions of spin-phonon relaxation times in a wide range of materials, providing microscopic insight into spin relaxation and guiding the development of spin-based quantum technologies.
\end{abstract} 
\maketitle
\begin{figure}[t]
\centering 
\includegraphics[width=0.95\columnwidth]{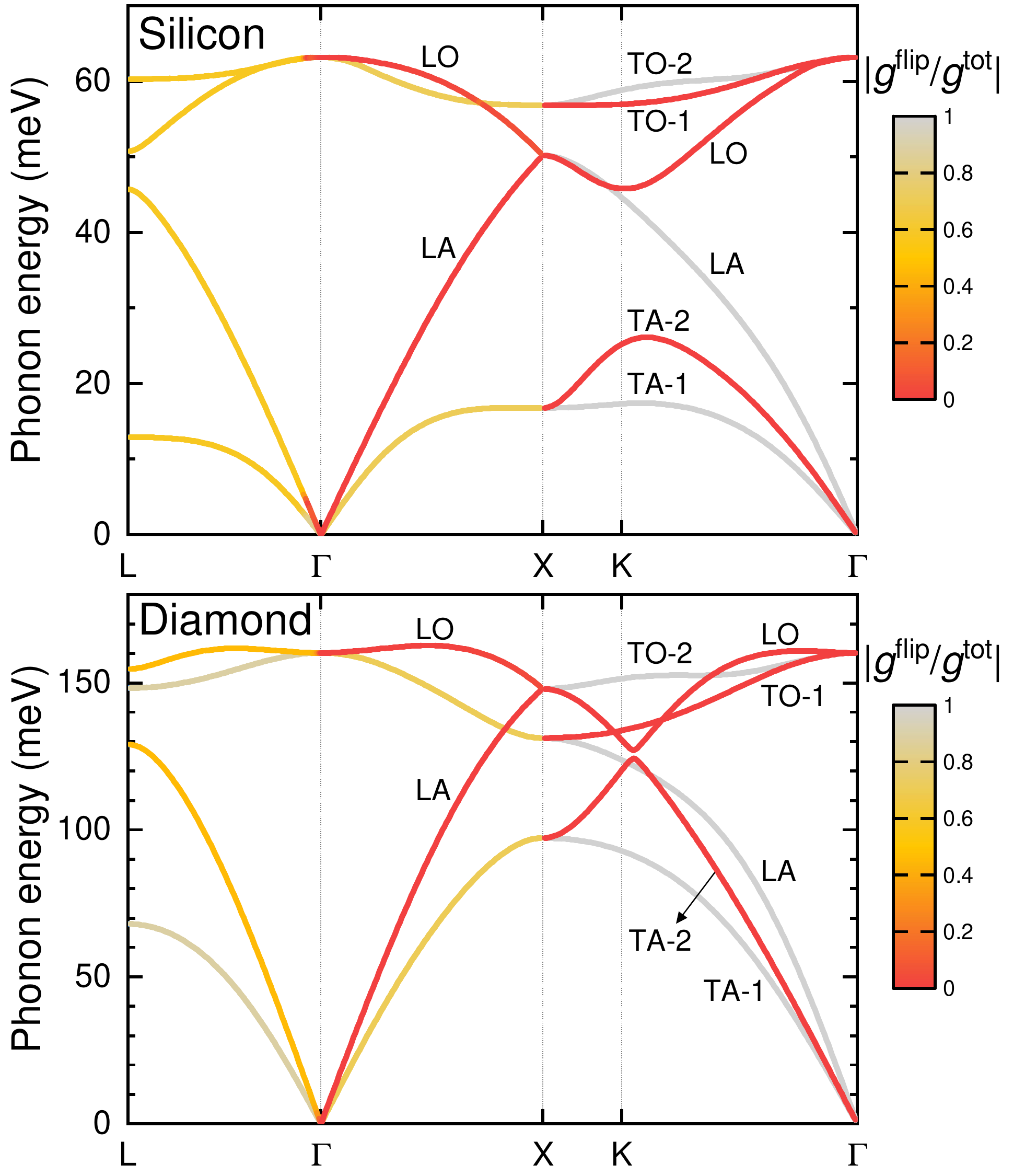}
\caption{Phonon dispersions in silicon and diamond, overlaid with a color map of the ratio $\abs{g_{\nu}^\text{flip}(\bm{q})/g_{\nu}^\text{tot}(\bm{q}) }$
between the spin-flip and the momentum-scattering $e$-ph matrix elements. The two matrix elements differ by orders of magnitude for the branches shown in red. 
The data shown are the square root of the gauge-invariant trace of $\abs{g}^2$ for a low-energy spin-degenerate conduction band~\cite{supp_mat}. 
The initial electron momentum is set to the $\Gamma$ point and we plot the ratio for phonon wave vectors $\bm{q}$ along a high-symmetry BZ line. 
}\label{fig:ephmats}
\end{figure}

%
\textit{\textbf{Introduction.}} 
Spin relaxation in centrosymmetric crystals primarily occurs through the Elliott-Yafet (EY) mechanism~\cite{elliottTheory1954,yafetFactors1963}, 
in which spin decoherence can be mediated by electron-phonon ($e$-ph) or electron-defect interactions. 
Phonons typically dominate EY spin relaxation near room temperature, and often limit the performance of spin-based devices 
in spintronics~\cite{balasubramanianNanoscale2008,jansenSilicon2012,hanGraphene2014} and quantum technologies~\cite{miCoherent2018,atatureMaterial2018,watsonProgrammable2018,widmann2015coherent}.
Recent advances are pushing spin manipulation to new frontiers~\cite{pesinSpintronics2012,mooreBirth2010,smejkalTopological2018,jungwirthMultiple2018}, 
so understanding in detail how electron spins interact with phonons is important for both technological and fundamental reasons.\\
\indent
%
%
Accurately predicting spin-phonon relaxation processes remains an open problem, particularly due to the challenge of quantifying spin-flip $e$-ph interactions~\cite{yafetFactors1963}. 
Calculations of EY spin relaxation have mainly relied on empirical models~\cite{chengTheory2010,kissElliottYafet2016} and symmetry analysis~\cite{tangElectron2012,songAnalysis2012,liIntrinsic2012,songTransport2013}, yet 
these approaches are laborious even for simple materials and not geared toward quantitative predictions. 
Attempts have also been made to study spin relaxation from first principles by assuming a direct proportionality between spin-flip and momentum-scattering $e$-ph interactions~\cite{restrepoFull2012}, 
or between spin-flip and momentum relaxation times~\cite{kurpasSpinorbit2016}. However, these assumptions hold only for simple model potentials~\cite{chazalvielSpin1975,fishmanSpin1977,fabianSpin1998} as   
the spin-flip and momentum-scattering processes can differ greatly depending on the electronic wave function, spin texture and phonon perturbation~\cite{ReviewRMP, songTransport2013}.\\
\indent
%
%
Recently developed first-principles methods for computing $e$-ph interactions and relaxation times~\cite{bernardiFirstprinciples2016} are promising for studying EY spin-phonon relaxation. 
Their typical workflow~\cite{agapitoInitio2018} involves density functional theory (DFT) calculations of the ground state and electronic band structure, 
combined with density functional perturbation theory (DFPT)~\cite{baroniPhonons2001} to compute the phonon dispersions and $e$-ph perturbation potentials,  
followed by interpolation of the $e$-ph coupling matrix elements to fine Brillouin zone (BZ) grids.
%
%
However, this workflow cannot be applied as is to investigate spin-flip $e$-ph interactions because the spin information is lost when one computes the e-ph matrix elements. 
For example, the electronic states in centrosymmetric crystals are at least two-fold degenerate, and their spin points in an arbitrary direction due to the freedom in describing the degenerate subspace. 
Computing spin-flip processes \textit{ab initio}, especially in the presence of spin-orbit coupling (SOC) and spinor wave functions, remains an open challenge.\\
\indent
%
Here we present a first-principles method for computing the spin-flip $e$-ph coupling matrix elements and the $T_1$ spin-phonon relaxation times (SRTs). 
Our approach assumes no relationship between the matrix elements for spin-flip and momentum scattering, 
and treats spinor wave functions and SOC through fully-relativistic DFT and DFPT calculations~\cite{theurichSelfconsistent2001}. 
These advances enable accurate calculations of SRTs and shed light on microscopic details of spin-phonon interactions.
We apply our method to investigate SRTs in two key materials for spintronic and quantum technologies, silicon and diamond. 
Our predicted SRTs in silicon are in excellent agreement with experiment between 50$-$300~K, while in diamond, where SRT measurements are missing, 
we predict intrinsic-limit SRTs of roughly 0.5~ms at 77~K and 2~$\mu$s at 300~K. 
In both materials, we find that spin-flip and momentum-scattering $e$-ph interactions differ widely and are not directly proportional, 
and the temperature dependence of the spin-flip and momentum relaxation times also differ greatly. 
Our work demonstrates a precise first-principles approach for computing SRTs, highlighting the limits of widely used simplified analyses and opening new avenues for microscopic understanding of spin dynamics.\\
\indent
%
%
\textit{\textbf{Spin-flip interactions.}} 
In centrosymmetric materials, the Bloch states with band index $n$ and crystal momentum $\bm{k}$ can be decomposed into \textit{effective} up and down spin states, 
denoted as $\Uparrow$ and $\Downarrow$, which diagonalize the $\hat{S}_z$ operator in the Kramers degenerate subspace~\cite{yafetFactors1963,elliottTheory1954,pientkaGauge2012}:
\begin{equation}
\begin{split}
&\braket{n\bm{k}{\Uparrow} | \hat{S}_z | {n\bm{k}}{\Uparrow}} = - \braket{n\bm{k}{\Downarrow} | \hat{S}_z | {n\bm{k}}{\Downarrow}}, \\
&\braket{n\bm{k}{\Downarrow} | \hat{S}_z | {n\bm{k}}{\Uparrow}}=0.
\end{split}
\end{equation}
%
%
The key ingredients for computing the SRTs are the spin-flip $e$-ph matrix elements~\cite{yafetFactors1963}, 
\begin{equation}\label{eq:gflip}
g_{mn\nu}^{\text{flip}}(\bm{k},\bm{q})= \braket{{m\bm{k}+\bm{q}}{\Downarrow} | \Delta \hat{V}_{\nu \bm{q}} | {n\bm{k}}{\Uparrow}} ,
\end{equation}
which quantify the probability amplitude to scatter from an initial Bloch state $\ket{n \bm{k} {\Uparrow} }$ to a final state $\ket{m \bm{k}+\bm{q} {\Downarrow} }$ 
with opposite effective spin, by emitting or absorbing a phonon with mode index $\nu$ and wave vector $\bm{q}$ due to the Kohn-Sham potential perturbation 
$\Delta \hat{V}_{\nu \bm{q}}$~\cite{bernardiFirstprinciples2016}, which is a $2\times2$ matrix in spin space in the presence of SOC.\\
\indent
%
%
To compute the SRTs, we obtain the effective spin states and from them the spin-flip $e$-ph matrix elements $g_{mn\nu}^{\text{flip}}(\bm{k},\bm{q})$ on fine BZ grids. 
We calculate the effective spin states from the spin matrix $S(\bm{k})$, which provides a matrix representation of the spin operator $\hat{S}_z$ in the wave function basis~\cite{mostofiUpdated2014}, 
$S_{ms',ns}(\bm{k})=\braket{m\bm{k} s'| \hat{S}_z | n\bm{k}s}$, where $s$ and $s'$ denote the spin. 
We diagonalize separately each degenerate subspace in the spin matrix at each $\bm{k}$-point, 
obtaining the unitary matrices $D_{\bm{k}}$ that make each of the subspaces in $D_{\bm{k}} {S}({\bm{k}})  D_{\bm{k}}^\dagger$ diagonal, with eigenvalues equal to the effective spin 
~\footnote{When only the two-fold degeneracy due to time-reversal plus inversion symmetry is present, the diagonal elements of $S(\bm{k})$ naturally determine the effective spin value.
For states with additional degeneracies, $D_{\bm{k}}$ diagonalizes the degenerate subspace, giving multiple pairs of states with opposite effective spin.}. 
The spin-flip $e$-ph matrix elements, $g_{mn\nu}^{\text{flip}}(\bm{k},\bm{q})$, are then computed using Eq.~(\ref{eq:gflip}) for all pairs of states with opposite effective spin.\\
\indent
%
%
\textit{\textbf{Interpolation.}} 
Since DFPT calculations of $\Delta \hat{V}_{\nu \bm{q}}$ on the fine BZ grids needed to converge the SRTs are prohibitively expensive, 
we interpolate the spin-flip $e$-ph matrix elements and spin matrices using Wannier functions~\cite{marzariMaximally2012,wangInitio2006,yatesSpectral2007}. 
To obtain $g_{mn\nu}^{\text{flip}}(\bm{k}',\bm{q}')$ at a desired pair of $\bm{k}'$ and $\bm{q}'$ points in the BZ, 
we first apply the usual Wannier interpolation workflow~\cite{agapitoInitio2018,giustinoElectronphonon2007} to obtain the $e$-ph matrix elements $g_{mn\nu}^{s s'}(\bm{k}',\bm{q}')$ between states with arbitrary spins $s$ and $s'$.  
The $e$-ph matrix elements $g_{mn\nu}^{\sigma \sigma'}(\bm{k}',\bm{q}')$ coupling states with effective spins $\sigma$ and $\sigma'$ are then computed using the unitary matrix $D_{\bm{k}'}$ 
(the latter is obtained from Wannier interpolation of the spin matrix~\cite{mostofiUpdated2014}):
\begin{equation}\label{eq:spinflipg}
g_{mn\nu}^{\sigma \sigma'}(\bm{k}',\bm{q}')= \left[D_{\bm{k}'+\bm{q}'}^{~} \right]_{m\sigma,ms} \left[ g_{mn\nu}^{s s'}(\bm{k}',\bm{q}') \right]\left[ D_{\bm{k}'}^\dagger \right]_{ns',n\sigma'}.
\end{equation}
The spin-flip $e$-ph matrix elements are finally computed between all pairs of electronic states with opposite sign of the effective spin. 
Our interpolation scheme can accurately reproduce spin-flip $e$-ph matrix elements obtained by combining effective spin states with perturbation potentials computed directly with DFPT 
(see the Supplemental Material~\cite{supp_mat}), thus enabling precise calculations of SRTs. 
\\
\indent
%
%
\textit{\textbf{Spin relaxation times.}} 
The band- and $\bm{k}$-dependent spin-flip $e$-ph relaxation times, $\tau^{\text{flip}}_{n\bm{k}}$, are computed using lowest-order perturbation theory~\cite{yafetFactors1963},
\begin{equation}\label{eq:taunkspin}
\begin{split}
\frac{1}{\tau^{\text{flip}}_{n\bm{k}}}=&\frac{4\pi}{\hbar}\sum_{m \nu \bm{q}}\abs{g_{mn\nu}^{\text{flip}}(\bm{k},\bm{q})}^2\\
&[(N_{\nu \bm{q}}+1-f_{m\bm{k+q}})\delta(\varepsilon_{n\bm{k}}-\varepsilon_{m\bm{k+q}}-\hbar\omega_{\nu\bm{q}}) \\
&~+ (N_{\nu \bm{q}}+f_{m\bm{k+q}})\delta(\varepsilon_{n\bm{k}}-\varepsilon_{m\bm{k+q}} +\hbar\omega_{\nu\bm{q}})],
\end{split}
\end{equation}
where  $\varepsilon_{n\bm{k}}$ and $\hbar\omega_{\nu \bm{q}}$ are the electron and phonon energies, respectively, and $f_{n\bm{k}}$ and $N_{\nu \bm{q}}$ the corresponding temperature-dependent occupations. 
Converging the BZ sum in Eq.~(\ref{eq:taunkspin}) is challenging, especially since the spin-flip $e$-ph matrix elements $g_{mn\nu}^{\text{flip}}(\bm{k},\bm{q})$ vary by several orders of magnitude throughout the BZ. 
We develop an importance sampling method to efficiently converge $\tau^{\text{flip}}_{n\bm{k}}$
(see the Supplemental Material ~\cite{ [{See \href{http://link.aps.org/supplemental/...}{Supplemental Material} for the importance sampling 
approach used in Eq.~(\ref{eq:taunkspin}), the states chosen for the comparison in Fig.~\ref{fig:ephmats},  
additional comparison of spin-flip and momentum-scattering matrix elements, momentum-scattering processes in diamond, 
and convergence of the interpolation scheme with respect to the coarse grid size}] supp_mat}).\\
\indent
%
%
The temperature-dependent SRT, $\tau_s(T)$, is the main physical observable computed in this work. 
It is obtained as an ensemble average of the spin-flip relaxation times~\cite{yafetFactors1963} by tetrahedron integration~\cite{blochlImproved1994}:
\begin{equation}\label{eq:taus}
\tau_s(T) =
{\left< \frac{1}{\tau^{\text{flip}}_{n\bm{k}}}\right>}^{-1}_{T}
=\left(    
\frac{\displaystyle\sum_{n \bm{k}} \displaystyle{ \frac{1}{\tau^{\text{flip}}_{n\bm{k}}} \left( -\frac{d f_{n\bm{k}}}{dE}\right)d\bm{k}}}{\displaystyle\sum_{n \bm{k}} \displaystyle{ \left( -\frac{d f_{n\bm{k}}}{dE}\right)d\bm{k}}}
\right)^{-1}.
\end{equation}
%
\begin{figure*}[t]
\centering 
\includegraphics[width=0.9\textwidth]{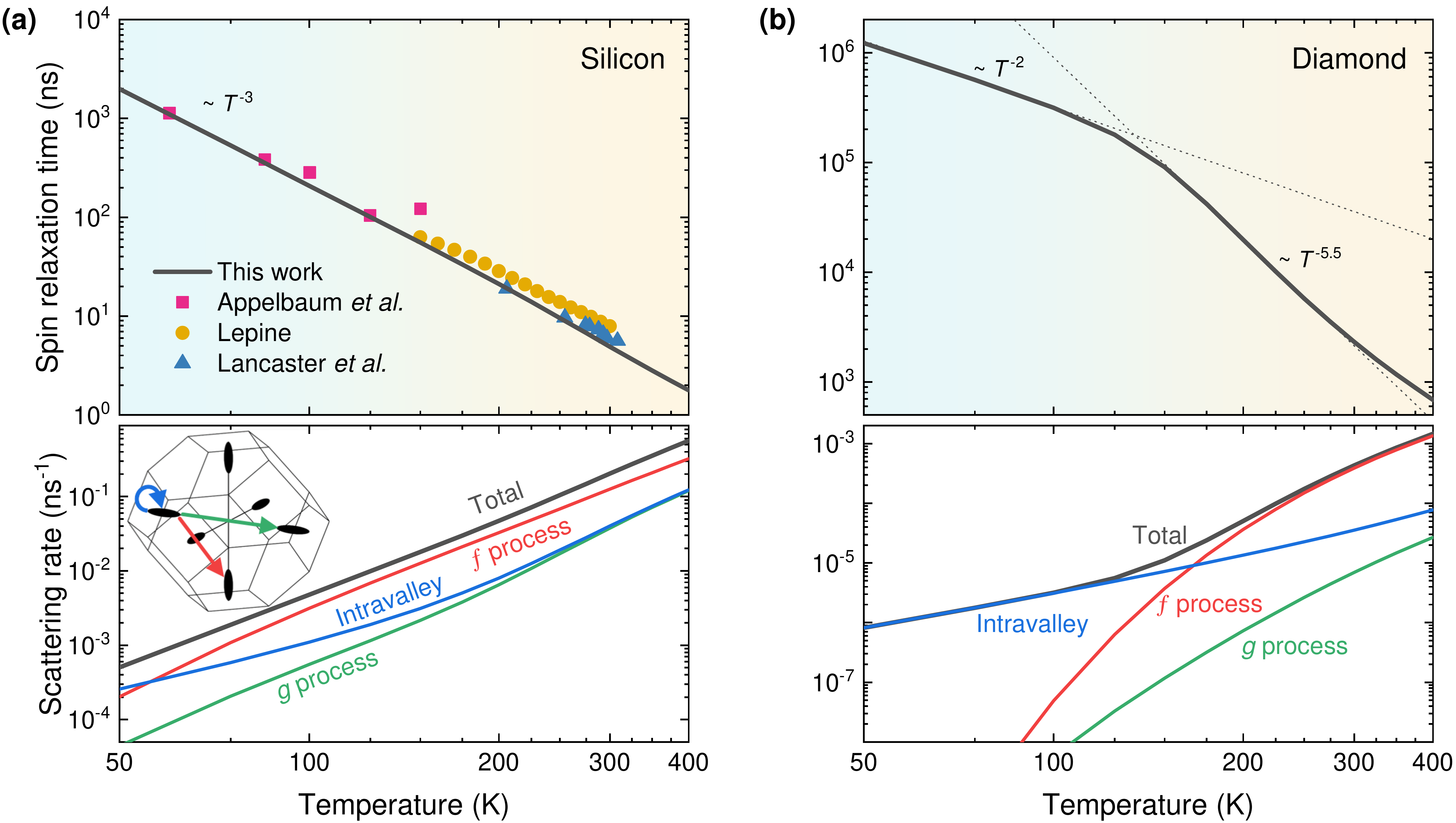}
\caption{Computed spin-phonon relaxation times as a function of temperature in (a) silicon and (b) diamond. The experimental data in (a) are taken from Refs.~\cite{appelbaumElectronic2007,*huangSpin2008,*appelbaumianIntroduction2011,lepineSpin1970,lancasterSpinlattice1964}. 
The lower panels show the process-resolved spin-flip $e$-ph scattering rates, defined as the inverse of $\tau_s$. 
Shown are the contributions from intravalley processes (blue line), $f$ processes (red line) and $g$ processes (green line), which add up to the total (gray line). 
The inset in (a) is a schematic of the intravalley and intervalley processes. 
}\label{fig:silicon_diamond}
\end{figure*}
%
%
\hspace{5pt}\textit{\textbf{Numerical methods.}} 
We apply our approach to investigate spin relaxation in silicon and diamond.  
We obtain their ground state and band structure using DFT with a plane-wave basis with the {\sc Quantum ESPRESSO} code~\cite{giannozziQUANTUM2009}. 
Briefly, we use relaxed lattice constants of 5.43~\text{\AA} for silicon and 3.56~\text{\AA} for diamond, together with a kinetic energy cutoff of 60~Ry for silicon and 120~Ry for diamond.  
We employ the PBEsol exchange-correlation functional~\cite{perdewRestoring2008} and fully-relativistic norm-conserving pseudopotentials~\cite{theurichSelfconsistent2001} 
from Pseudo Dojo~\cite{vansettenPseudoDojo2018}, which correctly include the SOC. 
We use DFPT~\cite{baroniPhonons2001} to compute the phonon dispersions and the perturbation potential, $\Delta \hat{V}_{\nu \bm{q}}$ in Eq.~(\ref{eq:gflip}), on coarse $\bm{q}$-point grids;  
our in-house developed {\sc perturbo} code~\footnote{The code employed in this work will be released in the future at \url{http://perturbo.caltech.edu}} 
is employed to compute the spin-dependent $e$-ph matrix elements on coarse BZ grids~\footnote{
The DFPT calculations are carried out on an $8\,\times\,8 \,\times\,8$ $\bm{q}$-point grid in diamond and a $10\,\times\,10 \,\times\,10$ $\bm{q}$-point grid in silicon. 
The spin-flip $e$-ph matrix elements are computed on $16\,\times\,16 \,\times\,16$ $\bm{k}$-point and $8\,\times\,8 \,\times\,8$ $\bm{q}$-point grids in diamond and 
$10\,\times\,10 \,\times\,10$ $\bm{k}$-point and $\bm{q}$-point grids in silicon.}. 
The DFPT calculations are done only in the irreducible $\bm{q}$-point grid, following which we extend the coarse-grid $e$-ph matrix elements to the full $\bm{q}$-point grid in {\sc perturbo} by rotating the spinor wave functions with $SU(2)$ matrices. 
The Wannier functions and spin matrices are computed with the {\sc Wannier90} code~\cite{mostofiUpdated2014} and employed in {\sc perturbo} to interpolate the spin-flip $e$-ph matrix elements 
on fine BZ grids with up to $200^3$ $\bm{k}$-points to converge the SRTs.  
The spin quantization axis is chosen as the $\left[001\right]$ direction.
\\
\indent
%
%
\textit{\textbf{Temperature-dependent SRTs.}} 
Figure~\ref{fig:silicon_diamond}(a) shows our calculated SRT as a function of temperature in silicon, 
which is in excellent agreement with experiments~\cite{appelbaumElectronic2007,*huangSpin2008,*appelbaumianIntroduction2011,lepineSpin1970,lancasterSpinlattice1964} at all temperatures between 50$-$300~K. 
For example, our calculated SRT at room-temperature is 4.9 ns, versus a 6.0 ns value measured by Lancaster \textit{et al.}~\cite{lancasterSpinlattice1964}~\footnote{We have verified that the results are nearly unchanged when using a different exchange-correlation functional. Using the same settings, the calculated SRT at 300~K is 4.8 ns with PBE and 4.5 ns with LDA}. 
%
%
The SRT in silicon exhibits an approximate $T^{-3}$ temperature dependence; 
to explain its origin, we analyze in Fig.~\ref{fig:silicon_diamond}(a) the contributions from the three valley-dependent scattering processes, 
including the intravalley and so-called $g$ and $f$ intervalley processes, which correspond to scattering between valleys along the same direction ($g$ processes) or along different directions ($f$ processes).
We find that the SRTs are comparable in magnitude for the three processes at all temperatures. 
The intravalley processes govern spin relaxation below 60~K, while $f$ intervalley scattering dominates at higher temperatures.\\
\indent
%
%
Due to its weak SOC and correspondingly long SRT, diamond is a promising material for spintronics and spin-based quantum technologies. 
However, SRT measurements have not yet been reported in diamond due to challenges related to spin injection~\cite{dohertyRoomTemperature2016}.
Figure~\ref{fig:silicon_diamond}(b) shows our computed SRT in diamond as a function of temperature.
We find SRTs of 540~$\mu$s at 77~K and 2.3~$\mu$s at 300~K; these values set an intrinsic limit due to phonons to the SRTs in diamond.  
%
%
The SRT exhibits a $T^{-2}$ temperature dependence below $\sim$170~K and a stronger $T^{-5.5}$ trend above 170~K.
This trend is in contrast with a previous prediction~\cite{restrepoFull2012} of a $T^{-5}$ temperature dependence 
throughout the entire temperature range and of an order-of-magnitude smaller SRT of 180 ns at room temperature. 
Ref.~\cite{restrepoFull2012} assumed a direct proportionality between the spin-flip and momentum-scattering $e$-ph matrix elements, 
but, as we show below, this assumption is in general incorrect and can lead to inaccurate phonon contributions to the SRT. 
%
We analyze the valley scattering processes in diamond in Fig.~\ref{fig:silicon_diamond}(b), and find that the intravalley processes dominate below 170~K, while the intervalley $f$ processes dominate above 170~K.\\ 
\indent
%
%
%
\textit{\textbf{Spin-flip versus momentum scattering.}} 
Our quantitative approach reveals stark differences between the spin-flip and the momentum-scattering interactions. 
Figure~\ref{fig:ephmats} compares the spin-flip coupling matrix elements, $\abs{g_{\nu}^\text{flip}(\bm{q})}$, with the spin-flip plus spin-conserving (i.e., momentum-scattering) $e$-ph matrix elements, $\abs{g_{\nu}^\text{tot}(\bm{q})}$, 
and resolves their ratio for different phonon modes. 
Depending on the phonon branch, we find that the spin-flip and momentum matrix elements can differ by several orders of magnitude, as we find for the longitudinal acoustic (LA) and longitudinal optical (LO) branches along $\Gamma-$X 
and for the LO and for specific transverse optical (TO-1) and transverse acoustic (TA-2) branches along X$-$K$-\Gamma$.  
For other phonon modes and BZ directions, the two quantities exhibit smaller $-$ yet quantitatively important $-$ differences. 
Only in specific cases the spin-flip and momentum-scattering interactions are nearly identical, as we find for the TO-2, TA-1 and LA branches along X$-$K$-\Gamma$. 
These trends are common to silicon and diamond. 
Analogous results are found when analyzing various initial and final electronic states~\cite{supp_mat}.\\ 
\indent
\begin{figure}[t]
\centering 
\includegraphics[width=1.0\columnwidth]{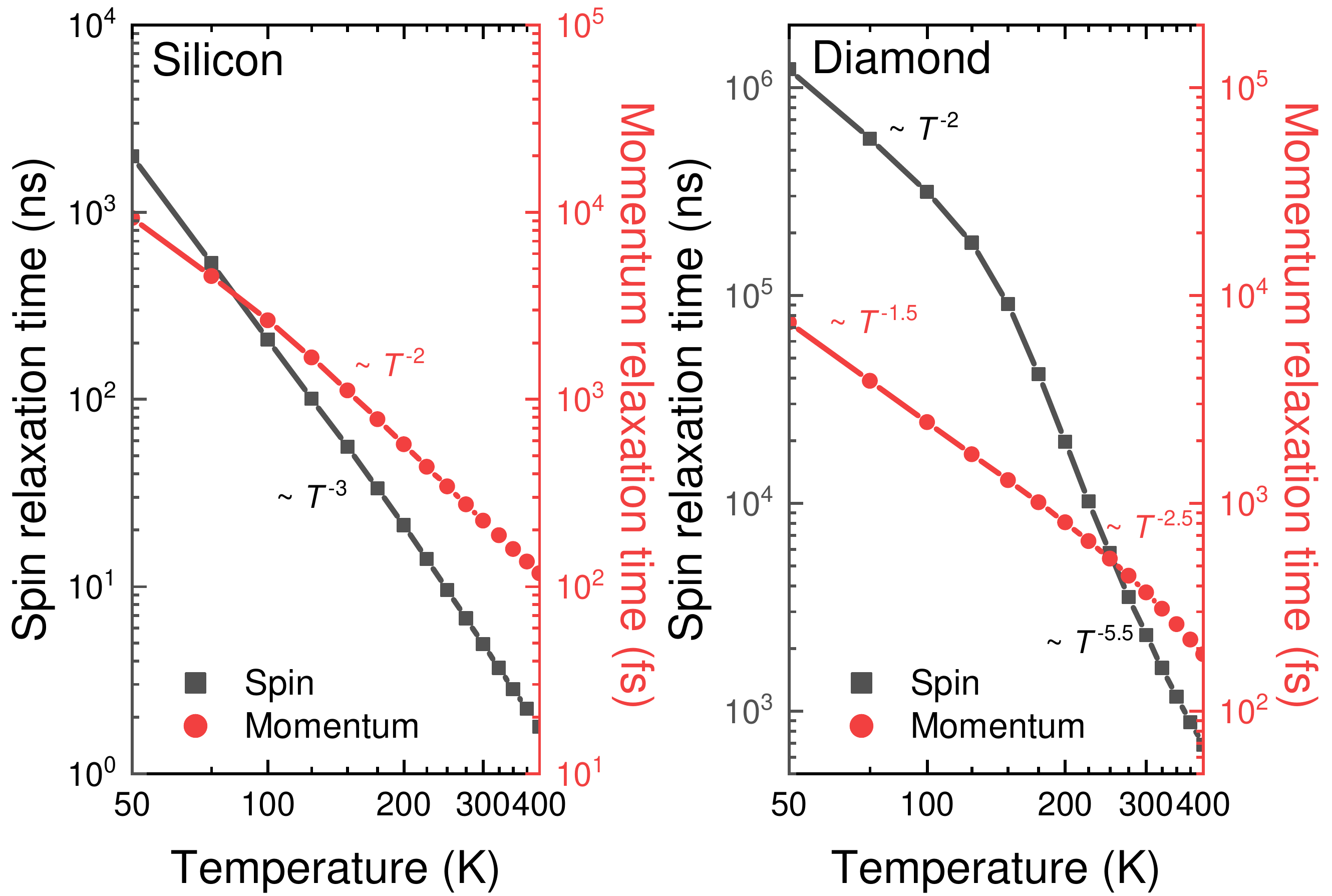}
\caption{Comparison between the temperature dependence of the SRT (gray squares) and the momentum relaxation time (red circles) in silicon and diamond. 
The labels give the exponent $n$ of the SRT temperature dependence, $T^{-n}$, separately for each of the spin and momentum relaxation times. 
Note that the SRTs are in ns units, and the momentum relaxation times in fs units.
}\label{fig:taus}
\end{figure}
%
%
Lastly, we compare the spin-phonon and momentum relaxation times. The momentum relaxation time $\tau_p$ is defined as the usual (spin-independent) $e$-ph relaxation time~\cite{bernardiFirstprinciples2016}, 
thermally averaged using Eq.~(\ref{eq:taus}) to make the comparison meaningful. 
The conventional wisdom is that spin and momentum relaxation times are directly proportional~\cite{elliottTheory1954,chazalvielSpin1975}, 
an assumption that has been widely used to analyze spin relaxation mechanisms in experimental data~\cite{hanSpin2011,zomerLongdistance2012,bandyopadhyayDominant2010,hanSpin2012,steckleinContactInduced2016,guiteTemperature2012}. 
%
%
Figure~\ref{fig:taus} shows the temperature dependent spin and momentum relaxation times in silicon and diamond. 
In silicon, the SRT follows a $T^{-3}$ temperature dependence, whereas the momentum relaxation time follows a $T^{-2}$ trend. 
In diamond, the SRT makes a sharp transition from a $T^{-2}$ trend at low temperature to a stronger $T^{-5.5}$ trend above 170~K.
In contrast, the momentum relaxation time exhibits a much weaker temperature dependence, roughly $T^{-1.5}$ at low temperature and $T^{-2.5}$ near room temperature.\\
\indent
%
%
There is no discernible direct proportionality between the spin and momentum relaxation times $-$ rather, they both exhibit an approximate $T^{-n}$ temperature dependence, but with different values  
of the exponent $n$ (see Fig.~\ref{fig:taus}). 
These differences originate from the different coupling strengths and phonon mode contributions, as we illustrate in Fig.~\ref{fig:ephmats}. For example, we find that for momentum scattering in diamond 
 the intravalley processes dominate over the entire temperature range up to 400~K, as opposed to just below 170~K as we show above for spin relaxation (see the Supplemental Material~\cite{supp_mat}).
We conclude that a reliable analysis of SRTs needs atomistic calculations that take into account the different nature of the spin-phonon and momentum-scattering $e$-ph interactions, 
using accurate spin-flip $e$-ph matrix elements.\\
\indent
%
%
\textit{\textbf{Discussion.}}  
Since SRT calculations involve a subtle interplay between spin-flip $e$-ph matrix elements and phonons and electronic states,  
the relative magnitude of the spin-phonon interactions for different phonon modes is of paramount importance for accurate predictions. 
Our results show that the widely used proportionality between spin and momentum relaxation times can be inaccurate, 
highlighting the need for atomistic details such as the electronic wave function, spin texture, phonon modes and their mode-dependent spin-flip interactions. 
When these microscopic details are captured, as we have shown above, one can predict the SRTs within $\sim$10$-$20\% of experiment over a wide temperature range, 
and predict which phonon modes govern spin relaxation. While computing $e$-ph interactions and carrier relaxation has become a main effort in 
first-principles calculations~\cite{bernardiInitio2014,zhouInitio2016,jhalaniUltrafast2017,leeCharge2018,zhouElectronPhonon2018}, 
SRT calculations are still in their infancy, and more work is needed to expand their scope beyond the EY mechanism discussed here. 
\\
\indent
\textit{\textbf{Conclusion.}} 
In summary, we have developed a quantitatively accurate approach for computing spin-flip $e$-ph interactions and SRTs due to the EY mechanism. 
The workflow proposed in this work is general $-$ it can be adapted to different perturbation potentials, including perturbations from defects~\cite{luEfficient2019}, 
through which one could study spin-flip and other defect-induced spin scattering processes.
Our approach can be applied broadly to study spin relaxation in materials for spintronics and magnetism, and in topological materials. 
It can also be extended to treat spin states localized at ions or defects. 
\begin{acknowledgments}
J. P. thanks Raffaello Bianco and I-Te Lu for fruitful discussions.
J. P. acknowledges support by the Korea Foundation for Advanced Studies.
This work was partially supported by the National Science Foundation under Grant No. CAREER-1750613, which provided for theory development, 
and by the Department of Energy under Grant No. de-sc0019166, which provided for numerical calculations and code development. 
\end{acknowledgments}
%
%
\bibliography{ref-sph}
\end{document}